\documentstyle[12pt,epsf,rotate]{article}

\newcommand{\eps}{\varepsilon}
\newcommand{\epe}{\varepsilon'/\varepsilon}
\newcommand{\mt}{m_{\rm t}}
\newcommand{\mc}{m_{\rm c}}
\newcommand{\ms}{m_{\rm s}}
\newcommand{\mw}{M_{\rm W}}
\newcommand{\gev}{\, {\rm GeV}}
\newcommand{\mev}{\, {\rm MeV}}
\newcommand{\bsi}{B_6^{(1/2)}}
\newcommand{\bei}{B_8^{(3/2)}}

\newcommand{\Lms}{\Lambda_{\overline{\rm MS}}}
\newcommand{\eqn}[1]{(\ref{#1})}

\begin{document}

\author{\\
{Andrzej J. BURAS${}^{1,2}$ and Markus E. LAUTENBACHER${}^{1}$}\\
{\small\sl ${}^{1}$ Physik Department, Technische Universit\"at
M\"unchen, D-85748 Garching, Germany.}\\
{\small\sl ${}^{2}$ Max-Planck-Institut f\"ur Physik
                    -- Werner-Heisenberg-Institut,}\\
{\small\sl F\"ohringer Ring 6, D-80805 M\"unchen, Germany.}
}

\date{}

\title{
{\large\sf
\rightline{MPI-Ph/93-60}
\rightline{TUM-T31-47/93}
\rightline{August 1993}
}
\vskip 1cm
{\LARGE\sf
An Analytic Formula and an Upper Bound for $\epe$ in the Standard
Model}\footnote{Supported by the German Bundesministerium f\"ur
Forschung und Technologie under contract 06 TM 732 and by the CEC
Science project SC1-CT91-0729.\\
{\scriptsize
email: {\tt buras@feynman.t30.physik.tu-muenchen.de} and
       {\tt lauten@feynman.t30.physik.tu-muenchen.de}} }
}

\maketitle
\thispagestyle{empty}

\begin{abstract}
\noindent
Using the idea of the penguin box expansion we find an analytic
expression for $\epe$ in the Standard Model as a function of $\mt$,
$\ms(\mc)$ and two non-perturbative parameters $\bsi$ and $\bei$.  This
formula includes next-to-leading QCD/QED short distance effects
calculated recently by means of the operator product expansion and
renormalization group techniques.  We also derive an analytic
expression for the upper bound on $\epe$ as a function of $|V_{\rm
cb}|$, $|V_{\rm ub}/V_{\rm cb}|$, $B_{\rm K}$ and other relevant
parameters.  Numerical examples of the bound are given.
\end{abstract}

\newpage
\setcounter{page}{1}

This year~\cite{burasetal:92d} we have analyzed the $CP$
violating ratio $\epe$ in the Standard Model including leading and
next-to-leading logarithmic contributions to the Wilson coefficient
functions of the relevant local operators. Another next-to-leading
order analysis of $\epe$ can be found in \cite{ciuchini:92}.

Imposing the constraints from the $CP$ conserving $K \rightarrow \pi \pi$
data on the hadronic matrix elements of these operators we have given
numerical results for $\epe$ as a function of $\Lms$, $\mt$ and two
non-perturbative parameters $\bsi$ and $\bei$ which cannot
be fixed by the $CP$ conserving data at present. These two parameters
are defined by
\begin{equation}
\langle Q_6(\mc) \rangle_0 \equiv \bsi \, \langle Q_6(\mc)
\rangle_0^{\rm (vac)}
\qquad
\langle Q_8(\mc) \rangle_2 \equiv \bei \, \langle Q_8(\mc)
\rangle_2^{\rm (vac)} \, ,
\label{eq:1}
\end{equation}
where
\begin{equation}
Q_{6} = \left( \bar s_{\alpha} \, d_{\beta}  \right)_{\rm V-A}
   \sum_{q} \left( \bar q_{\beta} \,  q_{\alpha} \right)_{\rm V+A}
\qquad
Q_{8} = \frac{3}{2} \left( \bar s_{\alpha} \, d_{\beta} \right)_{\rm
V-A}
         \sum_{q} e_{q} \left( \bar q_{\beta} \,  q_{\alpha}\right)_{\rm V+A}
\label{eq:2}
\end{equation}
are the dominant QCD and electroweak penguin operators, respectively.
The subscripts on the hadronic matrix elements denote the isospin of
the final $\pi\pi$-state.  The label ``vac'' stands for the vacuum
insertion estimate of the hadronic matrix elements in question for
which $\bsi=\bei=1$. The same result is found in the large $N$ limit
\cite{bardeen:87,burasgerard:87}. Also lattice calculations give
similar results $\bsi=1.0 \pm 0.2$ \cite{kilcup:91,sharpe:91} and $\bei
= 1.0 \pm 0.2$ \cite{kilcup:91,sharpe:91,bernardsoni:89,francoetal:89}.
We have demonstrated in \cite{burasetal:92d} that in QCD the parameters
$\bsi$ and $\bei$ depend only very weakly on the renormalization scale
$\mu$ when $\mu > 1\gev$ is considered. The $\mu$ dependence for the
matrix elements in \eqn{eq:1} is then given to an excellent accuracy by
$1/\ms^2(\mu)$ with $\ms(\mu)$ denoting the running strange quark
mass.

In the present letter we would like to cast the numerical results of
\cite{burasetal:92d} into an analytic formula which exhibits the
$\mt$-dependence of $\epe$ together with the dependence on $\ms$,
$\bsi$ and $\bei$. Such an analytic formula should be useful to those
phenomenologists and experimentalists who are not interested in getting
involved with the technicalities of ref.~\cite{burasetal:92d}.
Combining the analytic formula obtained here with the analytic lower
bound on $\mt$ derived recently by one of us \cite{buras:93} we will be
able to find an analytic upper bound on $\epe$.

In order to find an analytic expression for $\epe$ which exactly
reproduces the results of \cite{burasetal:92d} we use the idea of the
penguin-box expansion (PBE) suggested and developed in
\cite{buchallaetal:91}. This method allows to express flavour-changing
neutral current (FCNC) processes, in particular $\epe$, as linear
combinations of universal $\mt$-dependent functions resulting from
various penguin and box diagrams common to many FCNC processes. The
$\mt$-independent coefficients of these functions are $\mu$ and
renormalization scheme independent. They depend in the case at hand
only on $\Lms$, $\ms$, $\bsi$ and $\bei$.

The basic ideas of PBE have been discussed at length in
ref.~\cite{buchallaetal:91} and will not be repeated here. The simplest
method for transforming the results of ref.~\cite{burasetal:92d} into
the analytic formula given below is presented at the end of section~3
of ref.~\cite{buchallaetal:91}. In this letter we give only the final
formula for $\epe$.

The resulting analytic expression for $\epe$ is then given as follows
\begin{equation}
\epe = 10^{-4} \, \left[\frac{{\rm Im}\lambda_t}{1.7 \times 10^{-4}}\right]
\, F(x_t) \, ,
\label{eq:3}
\end{equation}
where
\begin{equation}
F(x_t) =
P_0 + P_X \, X(x_t) + P_Y \, Y(x_t) + P_Z \, Z(x_t) + P_E \, E(x_t) \, .
\label{eq:3b}
\end{equation}
Next
\begin{equation}
{\rm Im}\lambda_t = |V_{\rm ub}| \, |V_{\rm cb}| \, \sin \delta =
\eta \, \lambda^5 \, A^2
\label{eq:4}
\end{equation}
in the standard parameterization of the CKM matrix
\cite{particledata:92} and in the Wolfenstein parameterization
\cite{wolfenstein:83}, respectively. Here
\begin{equation}
\lambda = |V_{\rm us}| = 0.22
\qquad \qquad
|V_{\rm cb}| = A \, \lambda^2
\label{eq:4b}
\end{equation}
and
\begin{equation}
\eta = R_{\rm b} \, \sin\delta
\qquad \qquad
R_{\rm b} = \frac{1}{\lambda} \, \left| \frac{V_{\rm ub}}{V_{\rm cb}} \right|
\, .
\label{eq:4c}
\end{equation}

The basic $\mt$-dependent functions are given by
\begin{eqnarray}
X(x_t) &=& \frac{x_t}{8} \,
\left[ \frac{x_t+2}{x_t-1} + \frac{3 x_t -6}{(x_t-1)^2} \ln x_t \right]
\nonumber \\
Y(x_t) &=& \frac{x_t}{8} \,
\left[ \frac{x_t-4}{x_t-1} + \frac{3 x_t}{(x_t-1)^2} \ln x_t \right]
\label{eq:5} \\
Z(x_t) &=&
-\frac{1}{9} \ln x_t +
\frac{18 x_t^4 - 163 x_t^3 + 259 x_t^2 - 108 x_t}{144 (x_t-1)^3} +
\nonumber \\
 & &
\frac{32 x_t^4 - 38  x_t^3 - 15  x_t^2 +  18 x_t}{72  (x_t-1)^4} \ln x_t
\nonumber \\
E(x_t) &=&
-\frac{2}{3} \ln x_t +
\frac{x_t^2 (15 - 16 x_t + 4 x_t^2)}{6 (1-x_t)^4} \ln x_t +
\frac{x_t (18 - 11 x_t - x_t^2)}{12 (1-x_t)^3} \nonumber
\end{eqnarray}
with $x_t = \mt^2/\mw^2$. In the range $100\gev \le \mt \le 300\gev$
these functions can be approximated to better than 3\% accuracy by the
following expressions \cite{buchallaetal:91}
\begin{eqnarray}
X(x_t) = 0.650 \, x_t^{0.59} &\qquad&
Y(x_t) = 0.315 \, x_t^{0.78} \label{eq:6} \\
Z(x_t) = 0.175 \, x_t^{0.93} &\qquad&
E(x_t) = 0.570 \, x_t^{-0.51} \nonumber \, .
\end{eqnarray}
The coefficients $P_i$ contain the physics below the $\mw$-scale. They
are given in terms of $\bsi \equiv \bsi(\mc)$, $\bei \equiv \bei(\mc)$
and $\ms(\mc)$ as follows
\begin{equation}
P_i = r_i^{(0)} + \, \left[ \frac{150\mev}{\ms(\mc)} \right]^2 \,
\left( r_i^{(6)} \, \bsi + r_i^{(8)} \, \bei \right) \, .
\label{eq:7}
\end{equation}

The $P_i$ are $\mu$-independent and renormalization scheme
independent.  They depend however on $\Lms \equiv \Lms^{(4)}$. In the
table below we give the numerical values of $r_i^{(0)}$, $r_i^{(6)}$
and $r_i^{(8)}$ for different values of $\Lms$ at $\mu=\mc=1.4\gev$ and
the NDR renormalization scheme.  The coefficients $r_i^{(0)}$,
$r_i^{(6)}$ and $r_i^{(8)}$ do not depend on $\ms(\mc)$ as this
dependence has been factored out. $r_i^{(0)}$ does, however, depend on
the particular choice for the parameter $B_2^{(1/2)}$ in the
parameterization of the matrix element $\langle Q_2 \rangle_0$. The
values given in the table correspond to the central value for
$B_2^{(1/2)}=6.7 \pm 0.9$ as determined from the $CP$ conserving data
in ref.~\cite{burasetal:92d}. Variation of $B_2^{(1/2)}$ in the full
allowed range introduces an uncertainty of at most 18\% in the
$r_i^{(0)}$ column of the table.  Since the parameters $r_i^{(0)}$ give
only subdominant contributions to $\epe$ keeping $B_2^{(1/2)}$ and
$r_i^{(0)}$ at their central values is a very good approximation.

For different $\mu$ and renormalization schemes the numerical values in
the table change without modifying the values of the $P_i$'s as it
should be. To this end also $\bsi$ and $\bei$ have to be modified as they
depend on the renormalization scheme and albeit weakly on $\mu$.

\begin{center}
$\Delta S=1$ PBE coefficients for various $\Lms$.

\begin{tabular}{|c||c|c|c||c|c|c|}
\hline
& \multicolumn{3}{c||}{$\Lms=0.20\gev$} &
  \multicolumn{3}{c| }{$\Lms=0.25\gev$} \\
\hline
$i$ & $r_i^{(0)}$ & $r_i^{(6)}$ & $r_i^{(8)}$ & $r_i^{(0)}$ & $r_i^{(6)}$ &
$r_i^{(8)}$ \\
\hline
\hline
0 & $-$4.323 & 9.388 & 2.109 & $-$4.406 & 10.660 & 1.957 \\
\hline
$X$ & 0.988 & 0.016 & 0 & 0.958 & 0.019 & 0 \\
\hline
$Y$ & 0.763 & 0.078 & 0 & 0.729 & 0.086 & 0 \\
\hline
$Z$ & 0.235 & $-$0.014 & $-$10.072 & 0.297 & $-$0.015 & $-$10.899 \\
\hline
$E$ & 0.359 & $-$1.207 & 0.411 & 0.339 & $-$1.327 & 0.462 \\
\hline
\multicolumn{7}{c}{} \\
\hline
& \multicolumn{3}{c||}{$\Lms=0.30\gev$} &
  \multicolumn{3}{c| }{$\Lms=0.35\gev$} \\
\hline
$i$ & $r_i^{(0)}$ & $r_i^{(6)}$ & $r_i^{(8)}$ & $r_i^{(0)}$ & $r_i^{(6)}$ &
$r_i^{(8)}$ \\
\hline
\hline
0 & $-$4.484 & 12.038 & 1.792 & $-$4.557 & 13.573 & 1.609 \\
\hline
$X$ & 0.933 & 0.022 & 0 & 0.910 & 0.026 & 0 \\
\hline
$Y$ & 0.701 & 0.094 & 0 & 0.676 & 0.103 & 0 \\
\hline
$Z$ & 0.361 & $-$0.017 & $-$11.779 & 0.427 & $-$0.019 & $-$12.737 \\
\hline
$E$ & 0.319 & $-$1.449 & 0.515 & 0.299 & $-$1.579 & 0.573 \\
\hline
\end{tabular}
\end{center}

The inspection of the table shows that the terms involving $r_0^{(6)}$
and $r_Z^{(8)}$ dominate the ratio $\epe$. The function $Z(x_t)$
represents a gauge invariant combination of $Z^0$- and
$\gamma$-penguins \cite{buchallaetal:91} which increases rapidly with
$\mt$ and due to $r_Z^{(8)} < 0$ suppresses $\epe$ strongly for large
$\mt$ \cite{flynn:89,buchallaetal:90}. The term $r_0^{(0)}$ which also
plays some role represents the contributions of $(V-A) \otimes (V-A)$
QCD penguins. The positive contributions of $X(x_t)$ and $Y(x_t)$
represent to large extent the $(V-A) \otimes (V-A)$ electroweak
penguins.  Finally $E(x_t)$ describes the residual $\mt$-dependence of
the QCD-penguins which is only a very small correction to the full
expression for $\epe$.

Combining the formula \eqn{eq:3} with the analytic lower bound on $\mt$
from $\eps_{\rm K}$ derived recently in \cite{buras:93} we can find an
analytic upper bound on $\epe$. Indeed the lower bound of
\cite{buras:93} corresponds to $\sin\delta=1$ at which ${\rm Im}\lambda_t$
is maximal. Since $F(x_t)$ decreases with increasing $x_t$ we find
\begin{equation}
\left[\epe\right]_{\rm max} =
10^{-4} \, \left[\frac{R_{\rm b} \, \lambda^5 \, A^2}{1.7 \times
10^{-4}}\right] \, F((x_t)_{\rm min})
\label{eq:9}
\end{equation}
where according to \cite{buras:93}
\begin{equation}
(x_t)_{\rm min} =
\left[ \frac{1}{2\, A^2} \, \left( \frac{1}{A^2 \, B_{\rm K} \, R_{\rm
b}} - 1.2 \right) \right]^{1.316}
\label{eq:10}
\end{equation}
with  the renormalization group invariant $B_{\rm K}$ giving the size
of the hadronic matrix element $\langle \bar{K}^0 | (\bar{s} d)_{V-A}
(\bar{s} d)_{V-A} | K^0 \rangle$. Formula \eqn{eq:9} gives the maximal
value for $\epe$ in the Standard Model consistent with the measured
value of $\eps_{\rm K}$ as a function of $|V_{\rm cb}|$, $|V_{\rm
ub}/V_{\rm cb}|$, $B_{\rm K}$, $B_6^{(1/2)}$, $B_8^{(3/2)}$, $\ms$ and
$\Lms$. It should be noted that this formula does not involve $\mt$
directly as $\mt$ has been eliminated by means of the lower bound of
\cite{buras:93}. It is evident that the upper bound increases with
increasing $|V_{\rm cb}|$, $|V_{\rm ub}/V_{\rm cb}|$, $B_{\rm K}$,
$B_6^{(1/2)}$, $\Lms$ and with decreasing $\ms$ and $B_8^{(3/2)}$.

\begin{figure}[h]
\centerline{
\epsfysize=3.3in
\rotate[r]{
\epsffile{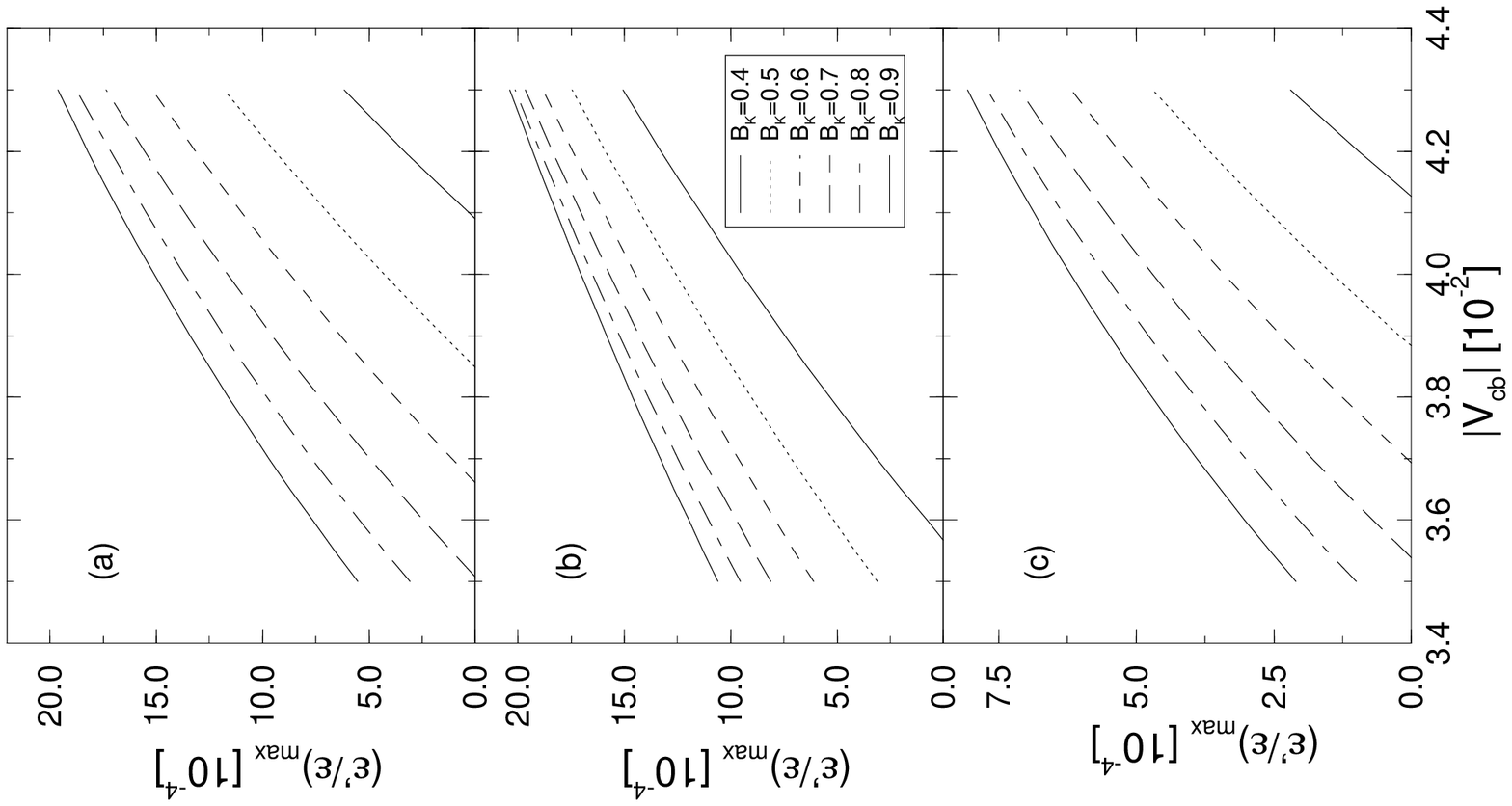}
}}
\caption[]{
$(\epe)_{\rm max}$ as a function of $|V_{\rm cb}|$ for $|V_{\rm
ub}/V_{\rm cb}| = 0.10$ and various choices of $B_{\rm K}$. The three
$(\epe)_{\rm max}$ plots correspond to hadronic parameters (a)
$B_6^{(1/2)}(\mc)=B_8^{(3/2)}(\mc)=2$, (b) $B_6^{(1/2)}(\mc)=2$,
$B_8^{(3/2)}(\mc)=1$ and (c) $B_6^{(1/2)}(\mc)=B_8^{(3/2)}(\mc)=1$,
respectively.  \label{fig:1}} \end{figure}

In fig.~\ref{fig:1} we show $(\epe)_{\rm max}$ as a function of
$|V_{\rm cb}|$ for different values of $B_{\rm K}$, three choices of
$(B_6^{(1/2)}, B_8^{(3/2)})$, $\Lms = 300\mev$, $\ms(m_c)=150\mev$ or
equivalently $\ms(1\gev)=175\mev$ and $|V_{\rm ub}/V_{\rm cb}| =
0.10$.
The range for $|V_{\rm cb}|$ has been chosen as in \cite{buras:93} in
accordance with the analyses of
\cite{neubert:9192,stone:92,ball:92,burdman:92} and
the increased $B$-meson life-time $\tau_{\rm B} = 1.49 \pm 0.03\,{\rm
ps}$ \cite{karlen:93}.

$B_6^{(1/2)}=B_8^{(3/2)}=1$ corresponds to the leading $1/N$
result or central values of lattice calculations. $B_6^{(1/2)}=2$,
$B_8^{(3/2)}=1$ is representative for the case advocated in
ref.~\cite{heinrichetal:92}. Finally, $B_6^{(1/2)}=B_8^{(3/2)}=2$ corresponds
effectively to the first case with a smaller value of $\ms$.  Comparing
fig.~\ref{fig:1} with the most recent messages from NA31 and E731
collaborations \cite{wagner:93,gibbons:93}

\begin{equation}
{\rm Re}(\epe) = \left\{
\begin{array}{ll}
(23 \pm 6.5)    \cdot 10^{-4} & {\rm NA31} \\
(7.4 \pm 6.0) \cdot 10^{-4} & {\rm E731}
\end{array} \right.
\label{eq:11}
\end{equation}
we observe that whereas the results of E731 are fully compatible with
the cases considered here, the NA31 data lie above the bounds of
fig.~\ref{fig:1} if $|V_{cb}| \le 0.040$. This is in particular the
case for $B_6^{(1/2)}=B_8^{(3/2)}=1$ which we advocate.

In fig.~\ref{fig:1} we only show $(\epe)_{\rm max} \ge 0$. We observe
that for $B_{\rm K} \le 0.5$ the upper bound is very low and for
$B_6^{(1/2)}=B_8^{(3/2)}=1$ and $|V_{\rm cb}| \le 0.039$ the ratio
$\epe$ becomes negative. This is simply the result of the high value of
$\mt$ required by $\eps_{\rm K}$ (see eq.~\eqn{eq:10}) when $B_{\rm K}$
and $|V_{\rm cb}|$ are low. For $B_{\rm K}=0.7 \pm 0.2$ obtained in the
$1/N$ and lattice calculations, $(\epe)_{\rm max}$ is in the ball park
of the E731 result. We note however that for $|V_{\rm cb}| \le 0.040$,
$B_{\rm K} \le 0.8$ and $B_6^{(1/2)}=B_8^{(3/2)}=1$ one has
$(\epe)_{\rm max} \le 5 \times 10^{-4}$. Since $|V_{\rm ub}/V_{\rm
cb}|$ is expected to be smaller than $0.10$
\cite{balletal:92,artuso:93} we should indeed be prepared for
$(\epe)_{\rm max} \approx {\rm few} \times 10^{-4}$ in the Standard
Model if $1/N$ calculations and lattice results are the full story.
This has been already emphasized in \cite{burasetal:92d,ciuchini:92}
but the bounds in fig.~\ref{fig:1}\,(c) make this point even stronger.
On the other hand we should keep in mind the strong dependence of the
upper bound \eqn{eq:9} on $B_{\rm K}$, $B_6^{(1/2)}$, $B_8^{(3/2)}$,
$\ms$, $|V_{\rm cb}|$ and $|V_{\rm ub}/V_{\rm cb}|$. Future
improvements on these parameters may bring surprises.

Finally it should be mentioned that below the upper bound ($\sin\delta
< 1$) the dependence of $\epe$ on $B_{\rm K}$, $|V_{\rm cb}|$ and
$|V_{\rm ub}/V_{\rm cb}|$ for fixed $\mt$ is more complicated and
different than in \eqn{eq:9}. $\epe$ increases with decreasing $B_{\rm
K}$ and $|V_{\rm cb}|$ at fixed $|V_{\rm ub}/V_{\rm cb}|$ until the
bound is reached. It increases (decreases) with decreasing $|V_{\rm
ub}/V_{\rm cb}|$ for $\pi/2 \le \delta \le \pi$ ($0 \le \delta \le
\pi/2$) at fixed $\mt$, $B_{\rm K}$ and $|V_{\rm cb}|$.

We hope that the analytic formula for $\epe$ and the analytic upper
bound on this important ratio presented in eqs.~\eqn{eq:3} and
\eqn{eq:9}, respectively, should facilitate the future phenomenological
analyses once the values for $|V_{\rm cb}|$, $|V_{\rm ub}/V_{\rm cb}|$,
$B_{\rm K}$, $B_6^{(1/2)}$, $B_8^{(3/2)}$, $\ms$, $\Lms$ and $\mt$ have
been improved and most importantly $\epe$ accurately measured.

\bigskip
\begin{center}
{\large\bf Acknowledgment}
\end{center}
\noindent
We gratefully acknowledge the most pleasant collaboration with Matthias
Jamin on $\epe$ in ref.~\cite{burasetal:92d} for which we supplemented
an analytic formula and an analytic upper bound in the present paper.

\newpage


{\small
\begin{thebibliography}{10}

\bibitem{burasetal:92d}
{\sc A.~J. Buras}, {\sc M.~Jamin}, and {\sc M.~E. Lautenbacher},
\newblock The Anatomy of $\varepsilon'/\varepsilon$ Beyond Leading Logarithms
  with Improved Hadronic Matrix Elements,
\newblock {\em Technical University Munich preprint, {\bf TUM-T31-35/92}, to
  appear in {\em Nucl.~Phys.}~{\bf B}} .

\bibitem{ciuchini:92}
{\sc M.~Ciuchini}, {\sc E.~Franco}, {\sc G.~Martinelli}, and {\sc L.~Reina},
\newblock {\em Phys. Lett.} {\bf B301} (1993) 263.

\bibitem{bardeen:87}
{\sc W.~A. Bardeen}, {\sc A.~J. Buras}, and {\sc J.-M. G\'{e}rard},
\newblock {\em Phys. Lett.} {\bf 180B} (1986) 133.

\bibitem{burasgerard:87}
{\sc A.~J. Buras} and {\sc J.-M. G\'{e}rard},
\newblock {\em Phys. Lett.} {\bf 192B} (1987) 156.

\bibitem{kilcup:91}
{\sc G.~W. Kilcup},
\newblock {\em Nucl. Phys.} {\bf B (Proc. Suppl.) 20} (1991) 417.

\bibitem{sharpe:91}
{\sc S.~R. Sharpe},
\newblock {\em Nucl. Phys.} {\bf B (Proc. Suppl.) 20} (1991) 429.

\bibitem{bernardsoni:89}
{\sc C.~Bernard} and {\sc A.~Soni},
\newblock {\em Nucl. Phys.} {\bf B (Proc. Suppl.) 9} (1989) 155.

\bibitem{francoetal:89}
{\sc E.~Franco} {\em et~al.},
\newblock {\em Nucl. Phys.} {\bf B 317} (1989) 63.

\bibitem{buras:93}
{\sc A.~J. Buras},
\newblock A 1993 Look at the Lower Bound on the Top Quark Mass from CP
  Violation,
\newblock {\em Technical University Munich preprint, {\bf TUM-T31-45/93};
  Max-Planck-Institut preprint, {\bf MPI-Ph/93-52}} .

\bibitem{buchallaetal:91}
{\sc G.~Buchalla}, {\sc A.~J. Buras}, and {\sc M.~K. Harlander},
\newblock {\em Nucl. Phys.} {\bf B349} (1991) 1.

\bibitem{particledata:92}
{\sc {Particle~Data~Group}},
\newblock Review of Particle Properties,
\newblock {\em Phys. Rev.} {\bf D45} (1992).

\bibitem{wolfenstein:83}
{\sc L.~Wolfenstein},
\newblock {\em Phys. Rev. Lett.} {\bf 51} (1983) 1841.

\bibitem{flynn:89}
{\sc J.~M. Flynn} and {\sc L.~Randall},
\newblock {\em Phys. Lett.} {\bf 224B} (1989) 221,
\newblock Erratum {\bf 235B} (1990) 412.

\bibitem{buchallaetal:90}
{\sc G.~Buchalla}, {\sc A.~J. Buras}, and {\sc M.~K. Harlander},
\newblock {\em Nucl. Phys.} {\bf B337} (1990) 313.

\bibitem{neubert:9192}
{\sc M.~Neubert},
\newblock {\em Phys. Lett. {\bf 264B}\,{\rm (1991)\,455}; Phys. Rev. {\bf
  D46}\,{\rm (1992)\,2212}} .

\bibitem{stone:92}
{\sc S.~L. Stone},
\newblock {\em $B$-Decays, {\rm p.~210, editor S. L. Stone, World Scientific,
  Singapore 1992}} .

\bibitem{ball:92}
{\sc P.~Ball},
\newblock {\em Phys. Lett.} {\bf 281B} (1992) 133.

\bibitem{burdman:92}
{\sc G.~Burdman},
\newblock {\em Phys. Lett.} {\bf 284B} (1992) 133.

\bibitem{karlen:93}
{\sc D.~Karlen},
\newblock B-Hadron Lifetimes,
\newblock {\em talk presented at the Heavy Flavours Conference, Montreal, July
  1993} .

\bibitem{heinrichetal:92}
{\sc J.~Heinrich}, {\sc E.~A. Paschos}, {\sc J.-M. Schwarz}, and {\sc Y.~L.
  Wu},
\newblock {\em Phys. Lett.} {\bf B279} (1992) 140.

\bibitem{wagner:93}
{\sc A.~{Wagner {\rm (NA31)}}},
\newblock {\em talk presented at the European Physics Symposium on High Energy
  Physics, Marseille, July 1993} .

\bibitem{gibbons:93}
{\sc L.~K. Gibbons} {\em et~al.},
\newblock {\em Phys. Rev. Lett.} {\bf 70} (1993) 1203.

\bibitem{balletal:92}
{\sc P.~Ball}, {\sc V.~M. Braun}, and {\sc H.~G. Dosch},
\newblock {\em Technical University Munich preprint, {\bf TUM-T31-31/92}, to
  appear in {\em Phys.~Rev.}~{\bf D}} .

\bibitem{artuso:93}
{\sc M.~Artuso},
\newblock {\em Syracuse University preprint, {\bf HEPSY-1-93}} .

\end{thebibliography}
}
\end{document}